\documentstyle[preprint,aps]{revtex}
\begin{document}
\preprint{DAMTP/R-97/16}
\draft
\tighten
\title{ABSORPTION OF FIXED SCALARS}

\author{Marika Taylor-Robinson \footnote{E-mail:
    mmt14@damtp.cam.ac.uk}}
\address{Department of Applied Mathematics and Theoretical Physics,
\\University of Cambridge, Silver St., Cambridge. CB3 9EW}
\date{\today}
\maketitle

\begin{abstract}
{We calculate the absorption rates of fixed scalars by near extremal
  five dimensional black holes carrying general one-brane and
  five-brane charges by semi-classical and D-brane methods. We find
  that the absorption cross-sections do not in general agree for
  either fixed scalar and we discuss possible explanations of the
  discrepancy.}

\end{abstract}
\narrowtext

\section{Introduction}
\noindent

In the past year there has been a great deal of interest in a certain
class of five-dimensional black holes with three gauge charges whose
properties may be reproduced by an effective string model based on
intersecting D-branes. Initially it was shown that such a model could
reproduce the Bekenstein-Hawking entropy of these black holes
\cite{VS}, \cite{HMS}, \cite{MS2}, \cite{JKM}, \cite{KT}
and more recently detailed comparisons of the emission rates have been
made \cite{CM}, \cite{DMW}, \cite{DM1}, \cite{DM2}, \cite{GK1},
\cite{GK}, \cite{MS}, \cite{KM}, \cite{MS1} \cite{Do}, \cite{HM}. 
Such calculations are straightforward both semi-classically
and in the string model for minimally coupled scalar fields. 

However
for minimally coupled scalars the absorption rate is not a sensitive
function of the moduli and energy; indeed for low energy neutral
scalar emission the absorption cross-section depends only on the
horizon area \cite{DGM}. A better test of the agreement between
semi-classical and string model calculations is provided by fixed
scalars \cite{FK} which are coupled to the gauge fields 
and whose absorption cross-sections behave very differently
to those of minimally coupled scalars. In \cite{CGKT}, \cite{KK1}, the 
string calculation for a particular fixed scalar, which is related to the
volume of the internal torus around which five branes wrap, was found
to agree with the semi-classical calculation in the case that the one
and five brane charges are the same. A particularly interesting
feature of this calculation is that the string cross-section involves
the tension of the effective string and thus one might expect
a calculation of the 
semi-classical cross-section for general charges to confirm that 
this tension is independent of the one-brane charge.

\bigskip

Although the string calculation was possible for all one and five
brane charges within the dilute gas region, the semi-classical
calculations were found to be technically difficult, owing to a
coupling between fixed scalar and gravitational perturbations 
when the charges were not the same. In this paper we discuss the
solution of the semi-classical field equations for the fixed scalars 
when the charges take general values. We calculate the absorption
cross-sections for the two fixed scalars related to the volume of the
internal torus and the scale of the effective string direction
respectively. We then approximate the cross-section for the latter
derived from the effective string action, and compare the results.

We find that for general values of the charges the cross-sections for
the fixed scalars calculated in the two different regimes do not
agree. There is a mixing between the two fixed scalars at the
semi-classical level when the charges are not equal, whereas there
does not appear to be any such mixing from the effective string point
of view. In the limit of low energy absorption
from a very near extremal state, the string cross-sections behave as
$T_{L}^5$, where $T_{L}$ is the temperature of the left movers, whereas
the semi-classical cross-sections behave as $\omega^2 T_{L}^3$, where 
$\omega$ is the frequency. This discrepancy derives from the presence
of chiral operators of dimensions $(3,1)$ and $(1,3)$ as well as an
operator of dimension $(2,2)$ in the string
interaction for one fixed scalar, whilst no such operators contribute
to the interactions of the other. 

The paper is organised as follows. In section II we discuss the semi-classical calculation of the absorption cross-sections for the
two fixed scalars, and in section III we calculate the functional
dependence of the absorption cross-sections implied by the string
theory effective action. We give our conclusions in section IV. 

\section{Semi-classical calculation of greybody factors}
\noindent

As usual we consider a class of five-dimensional black hole
representing the bound state of $n_{1}$ RR strings and $n_{5}$ RR
5-branes compactified on a 5-torus first discussed in \cite{VS},
\cite{HMS}.  These black holes can be regarded
as a solution to a truncation of type
IIB superstring effective action compactified on a 5-torus
\begin{eqnarray}
S_{5} = \frac{1}{2 \kappa_{5}^2} \int d^5x \sqrt{g} [R - (\partial
\phi)^2 - \frac{4}{3} (\partial \lambda)^2 - 4 (\partial \nu)^2
\label{act} \\
- \frac{1}{4} e^{\frac{8\lambda}{3}}F^{(K)2} - \frac{1}{4}e^{-
  \frac{4\lambda}{3} + 4 \nu} F^2 - \frac{1}{4} e^{-
  \frac{4\lambda}{3} - 4\nu} H^2] \nonumber
\end{eqnarray}
where $F^{(K)}_{\mu\nu}$ is the Kaluza-Klein vector field strength
corresponding to the string direction and $F_{\mu\nu}$, $H_{\mu\nu}$
are the ``electric'' and ``magnetic'' components of the field strength
of the RR two form. As discussed in \cite{CGKT}, the dilaton $\phi$ is
a decoupled scalar, whilst the scalars $\lambda$ and $\nu$ interact
with the gauge fields and are examples of fixed scalars \cite{FK}. The
scalar $\lambda$ is related to the scale of the Kaluza-Klein circle
whilst $\nu$ is related to the scale of the internal torus. 

Following \cite{CGKT}, we choose the five dimensional metric as
\begin{equation}
ds_{5}^{2} = - e^{2a(t,r)}dt^2 + e^{2b(t,r)} dr^2 + e^{2c(t,r)}
d\Omega_{3}^2, 
\end{equation} 
where we assume that the metric functions are angular independent
since we are interested in low energy scattering for which only $l=0$
components are significant. Now
the graviton equations of motion take the forms
\begin{equation}
R_{\mu\nu} = \frac{4}{3} \partial_{\mu}\lambda \partial_{\nu}\lambda
+ 4 \partial_{\mu} \nu \partial_{\nu} \nu + T^{M}_{\mu\nu},
\end{equation}
where the trace adjusted energy tensor for the gauge fields is 
\begin{eqnarray}
T^{M}_{\mu\nu} =
e^{\frac{8\lambda}{3}}(\frac{1}{2} F^{(K)}_{\mu\lambda}F^{(K)\lambda}_{\nu} 
- \frac{1}{12} F^{(K)2} g_{\mu\nu}) + 
e^{-\frac{4\lambda}{3} + 4 \nu} (\frac{1}{2}
F_{\mu\lambda}F^{\lambda}_{\nu} \\ 
- \frac{1}{12} F^{2} g_{\mu\nu}) + \frac{1}{4} e^{-\frac{4\lambda}{3} - 4\nu}  
(\frac{1}{2} H_{\mu\lambda}H^{\lambda}_{\nu} 
- \frac{1}{12} H^{2} g_{\mu\nu}), \nonumber
\end{eqnarray}
and we have omitted dilaton components, since we can choose the
background value of the dilaton to be constant and can set all
variations to vanish. 
Solving the equations of motion for the gauge fields we find that
\begin{eqnarray}
F^{(K)}_{tr} = 2Q_{K} e^{a+b-3c- \frac{8}{3} \lambda} \\
F_{tr} = 2Q e^{a+b-3c+\frac{4}{3} \lambda - 4 \nu} \\
H_{tr} = 2P e^{a+b-3c+\frac{4}{3} \lambda + 4 \nu}
\end{eqnarray}
where $Q_{K}$, $Q$ and $P$ are the gauge charges. We can then
eliminate these fields from the action, as in \cite{CGKT}, replacing
them with a potential dependent on the fixed scalars and the metric
function $c$ only. Thus the equation for the fixed scalar $\nu$ takes the
form
\begin{equation}
\partial_{t} (e^{-a+b+3c} \dot{\nu}) - \partial_{r}(
e^{a-b+3c} \nu') = e^{a+b-3c} [- P^2 e^{\frac{4\lambda}{3} + 4\nu} + Q^2 
e^{\frac{4\lambda}{3} - 4\nu}],
\end{equation}
whilst the equation for the fixed scalar $\lambda$ is
\begin{equation}
\partial_{t} ( e^{-a+b+3c} \dot{\lambda}) - \partial_{r}(
e^{a-b+3c} \lambda') = 
e^{a+b-3c} [2Q_{K}^{2} e^{\frac{8\lambda}{3}} -
P^2 e^{\frac{4\lambda}{3} + 4\nu} - Q^2 
e^{\frac{4\lambda}{3} - 4\nu}].
\end{equation}
We are interested in linearising these equations of motion about the
background metric 
\begin{equation}
e^{2a_{0}} = h {\mathcal H}^{-2/3}, e^{2b_{0}} = h^{-1} {\mathcal H}^{1/3},
e^{2c_{0}} = r^2 {\mathcal H}^{1/3},  
\end{equation}
in which the fixed scalars are given by
\begin{equation}
e^{2\lambda_{0}} = H_{r_{K}^2} (H_{r_{1}^2} H_{r_{5}^2})^{-1/2},
e^{4\nu_{0}} = H_{r_{1}^2} H_{r_{5}^2}^{-1}, 
\end{equation}
and the metric functions are defined as 
\begin{eqnarray}
h = (1 - \frac{r_{0}^2}{r^2} ) &,& H_{r_{i}^2} = 1 +
\frac{r_{i}^{2}}{r^2}, \\
r_{i}^{2} = \sqrt{Q_{i}^2 + \frac{1}{4}
  r_{0}^4} - \frac{1}{2} r_{0}^2 &,& 
{\mathcal H} = H_{r_{1}^2} H_{r_{5}^2} H_{r_{K}^2}, \nonumber 
\end{eqnarray}
where we have introduced the characteristic radii $r_{i}$ and $r_{0}^2$ is
an extremality parameter. In the extremal
limit, $r_{1}^2 = Q$, $r_{5}^2 = P$ and $r_{K}^2 = Q_{K}$. We shall be
interested in the dilute gas limit for which $r_{K} = r_{0} \sinh
\sigma$, and the effective temperatures of the left and right moving
modes on the effective string are related to the radial parameters as
\begin{equation}
T_{L} = \frac{1}{2\pi} ({\frac{r_{0}}{r_{1}r_{5}}}) e^{\sigma},
\hspace{5mm}
T_{R} = \frac{1}{2\pi} ({\frac{r_{0}}{r_{1}r_{5}}}) e^{-\sigma}.
\end{equation}
To obtain the linearised equations of motion, we let the scalar
functions take the form $f = f_{0} + \delta f$. We will use 
the residual gauge freedom to fix the metric on the sphere, i.e. $\delta
c =0$. Since the background is static, we can replace all coefficients
of time derivatives in the equations of motion with their background
values. For the radial-time graviton equation this produces the
simplified linearised equation
\begin{equation}
R_{tr} = \frac{4}{3} \lambda_{0}' \delta \dot{\lambda}
+ 4 \nu_{0}' \delta \dot{\nu},
\end{equation}
and from the form of the metric we can calculate the relevant component
of the Ricci tensor as $3 c_{0}' \delta \dot{b}$ so that
\begin{equation}
\delta b =  \frac{1}{3 c_{0}'} (\frac{4}{3} 
\lambda_{0}' \delta \lambda + 4 \nu_{0}' \delta \nu) \label{beq}
\end{equation}
where we have integrated with respect to the time coordinate.

It is then convenient to use the angular graviton equations of motion;
the angular components of the Ricci tensor depend only on the zeroth
and first derivatives of the perturbations, whilst the time and radial
components depend also on second derivatives. That is, 
\begin{equation}
\delta R_{\theta\theta} = g_{\theta\theta} e^{-2b_{0}} 
\lbrace c_{0}'(\delta b' - \delta a') + 2 (c_{0}'(a_{0}' - b_{0}' +
3c_{0}') + c_{0}'') \delta b \rbrace,
\end{equation} 
where $\theta$ is a coordinate on the sphere. From the graviton
equation of motion, however, we find that
\begin{equation}
\delta R_{\theta\theta} = \frac{1}{12} g_{\theta\theta} \delta [
e^{\frac{8}{3}\lambda} F^{(K)2} + e^{-\frac{4}{3}\lambda + 4\nu} F^{2}
+  e^{-\frac{4}{3}\lambda - 4\nu} H^{2}].
\end{equation}
We hence find a relation between the first derivatives of the
gravitational perturbations
\begin{eqnarray}
(\delta a' - \delta b') - f(r) \delta b &=&
\frac{2{\mathcal{H}}^{-2/3}}{3 r^6 h c_{0}'} [
\frac{8Q_{K}^2}{3}e^{-\frac{8}{3} \lambda_{0}} \delta \lambda 
- Q^2 e^{\frac{4}{3}\lambda_{0} - 4 \nu_{0}} (\frac{4}{3} \delta \lambda - 4
\delta \nu) \label{abeq} \\
&& - P^2 e^{\frac{4}{3} \lambda_{0}+ 4 \nu_{0}} ( \frac{4}{3}
\delta \lambda + 4 \delta \nu)], \nonumber
\end{eqnarray}
where the function $f(r)$ is given by
\begin{equation}
f(r) = 2 ( \frac{c_{0}''}{c_{0}'} + \frac{3}{r} + \frac{h'}{h})
\end{equation}
Given the two equations (\ref{beq}) and (\ref{abeq}) 
for the gravitational perturbations, we can 
substitute into the linearised equations for the fixed scalars
\begin{eqnarray}
e^{-a_{0}+b_{0}+3c_{0}} \delta \ddot{\nu} - \delta(\partial_{r}(
e^{a-b+3c_{0}} \delta \nu')) 
= \delta(e^{a+b-3c_{0}} [-P^2 e^{\frac{4\lambda}{3} + 4\nu} + Q^2 
e^{\frac{4\lambda}{3} - 4\nu}]), \\
e^{-a_{0}+b_{0}+3c_{0}} \delta \ddot{\lambda} - \delta(\partial_{r}(
e^{a-b+3c_{0}} \lambda')) = \delta \lbrace 
e^{a+b-3c_{0}} [2Q_{K}^{2} e^{\frac{8\lambda}{3}} -
P^2 e^{\frac{4\lambda}{3} + 4\nu} - Q^2 
e^{\frac{4\lambda}{3} - 4\nu}] \rbrace, 
\end{eqnarray}
and decouple the equations for the fixed scalars from those for the
gravitational perturbations. For modes of frequency $\omega$ such that
$\delta f = \tilde{f} e^{i\omega t}$, we find the following coupled
equations for the fixed scalars, 
\begin{equation}
[(h r^3 \partial_{r})^2 + \omega^2 r^6 {\mathcal H} + F_{\lambda}]
\tilde{\lambda} + 3 F(r) \tilde{\nu} = 0  
\end{equation}
\begin{equation}
[(h r^3 \partial_{r})^2 + \omega^2 r^6 {\mathcal H} + F_{\nu}]
\tilde{\nu} +  F(r) \tilde{\lambda} = 0, \nonumber 
\end{equation}
where the functions are given by
\begin{eqnarray}
F_{\lambda} &=& - \frac{8 h r^4}{(r_{1}^2 r_{5}^2 + r_{1}^2 r_{K}^2
+ r_{5}^2 r_{K}^2 + 2 (r_{1}^2 + r_{5}^2 + r_{K}^2) r^2 + 3 r^4)^2}
\lbrace r_{1}^4 r_{5}^4 + r_{1}^4 r_{K}^4 +r_{5}^4 r_{K}^4  \nonumber \\
&& + 2 r_{1}^2 r_{5}^2 r_{K}^2 (r_{1}^2 + r_{5}^2 + r_{K}^2) + 
((r_{1}^2 r_{5}^2 + 4 r_{K}^4) + r_{K}^2 (r_{1}^2 + r_{5}^2 )  
+ 
r_{K}^2 ( r_{1}^4 + r_{5}^4) \\
&& + 6 r_{1}^2 r_{5}^2 r_{K}^2 )r^2 
+ (r_{1}^4 + r_{5}^4 - r_{1}^2 r_{5}^{2} + 4 r_{K}^4 + 2 r_{K}^2
r_{1}^{2} + 2 r_{K}^2r_{5}^2 ) r^4 \rbrace, \nonumber 
\end{eqnarray}
and 
\begin{eqnarray}
F_{\nu} &=& - \frac{8 h r^4}{(r_{1}^2 r_{5}^2 + r_{1}^2 r_{K}^2
+ r_{5}^2 r_{K}^2 + 2 (r_{1}^2 + r_{5}^2 + r_{K}^2) r^2 + 3 r^4)^2}
\lbrace r_{1}^4 r_{5}^4 + r_{1}^4 r_{K}^4 +r_{5}^4 r_{K}^4  \nonumber \\
&& + 2 r_{1}^2 r_{5}^2 r_{K}^2 (r_{1}^2 + r_{5}^2 + r_{K}^2) +
3 (r_{1}^2 r_{5}^2 (r_{1}^2 + r_{5}^2 ) + (r_{1}^4 + r_{5}^4 ) r_{K}^2
+ 2 r_{1}^2 r_{5}^2 r_{K}^2 ) r^2 \\
&& + 3 ( r_{1}^4 + r_{5}^4 + r_{1}^2 r_{5}^2 ) r^4 \rbrace, \nonumber 
\end{eqnarray}
and 
\begin{equation}
F(r) = 8 h r^6 \frac{(r_{1}^2 - r_{5}^2)[r_{1}^2 r_{5}^2 + r_{1}^2 r_{K}^2
+ r_{5}^2 r_{K}^2 + (r_{1}^2 + r_{5}^2 + r_{K}^2) r^2]}{(r_{1}^2
r_{5}^2 + r_{1}^2 r_{K}^2
+ r_{5}^2 r_{K}^2 + 2 (r_{1}^2 + r_{5}^2 + r_{K}^2) r^2 + 3 r^4)^2}.
\end{equation}
For notational simplicity, we have assumed that the black hole is very
near extremal, so that $r_{0}^2 \ll r_{1}^ 2, r_{5}^2$, and we
accordingly retain only leading order terms. Note that when the 
one and five brane charges are equal the equation for the fixed scalar
$\tilde{\nu}$ reduces to 
\begin{equation}
[(h r^3 \partial_{r})^2 + \omega^2 r^6 {\mathcal H} - \frac{8 h r^4
  R^4}{(r^2 + R^2)^2} ( 1 + \frac{r_{0}^2}{R^2})] \tilde{\nu} = 0,
\end{equation}
where we have set $r_{1} = r_{5} \equiv R$; this is indeed the
equation obtained in \cite{CGKT}.  Note that we have restored
the full dependence on the non extremality parameter so that the
equation is exact. The corresponding equation
for the fixed scalar $\tilde{\lambda}$ reduces to 
\begin{equation}
[(h r^3 \partial_{r})^2 + \omega^2 r^6 {\mathcal H} - \frac{8 h r^4
  (R^2+ 2 r_{K}^2)^2}{(3r^2 + (R^2+ 2r_{K}^2))^2} 
( 1 + \frac{r_{0}^2}{(R^2 + 2 r_{K}^2)})] \tilde{\lambda} = 0,
\end{equation}
which in the dilute gas region $r_{K} \ll R$ differs from the equation
for the other fixed scalar by only a factor in the effective potential
term. 

Now the equations for the
fixed scalars are in general coupled, although it is apparent from the
form of $F(r)$ that the equations decouple when the one and five brane
charges are equal as was found in \cite{CGKT}. We can however find a
linear transformation of the fields which decouples the equations;
introducing two scalar fields $\phi_{a}$, $\phi_{b}$ the required
transformation is 
\begin{eqnarray}
\tilde{\lambda} &=& \sqrt{3} [\phi_{a} \cos{\psi} - \phi_{b} \sin{\psi}],
\\
\tilde{\nu} &=& \phi_{a}\sin{\psi} + \phi_{b} \cos{\psi}, \nonumber 
\end{eqnarray}
where the mixing angle $\psi$ is defined by the quadratic equation 
\begin{equation}
\tan^2{\psi} - \frac{2}{\sqrt{3}}(\frac{r_{1}^2 + r_{5}^2 - 2
  r_{K}^2}{r_{1}^2 - r_{5}^2}) \tan{\psi}  - 1 = 0,
\end{equation}
and hence we find that 
\begin{equation}
\sin^2 {\psi} = \frac{1}{2} \pm \frac{1}{4} \frac{r_{1}^2 + r_{5}^2 -
2 r_{K}^2 }{\sqrt{r_{1}^4 + r_{5}^4 + r_{K}^4 - r_{1}^2 r_{5}^2 
- r_{1}^2 r_{K}^2 - r_{5}^2 r_{K}^2}}. \label{hp}
\end{equation}
The transformed fields satisfy the equations
\begin{equation}
[(h r^3 \partial_{r})^2 + \omega^2 r^6 {\mathcal H} - 8
\frac{h r^4 r^4_{a,b}}{(r^2 + r_{a,b}^2)^2 } (1 +
\frac{r_{0}^2}{r_{a,b}^2})] \phi_{a,b} = 0, \label{up}
\end{equation}
where the effective radii $r_{a,b}$ are defined as 
\begin{eqnarray} 
r_{a}^2 = \frac{1}{3} [r_{1}^2 + r_{5}^2 + r_{K}^2 + \sqrt{r_{1}^4 +
  r_{5}^4 + r_{K}^4 - r_{1}^2 r_{5}^2 - r_{1}^2 r_{K}^2 - r_{5}^2
  r_{K}^2}], \\
r_{b}^2 = \frac{1}{3} [r_{1}^2 + r_{5}^2 + r_{K}^2 - \sqrt{r_{1}^4 +
  r_{5}^4 + r_{K}^4 - r_{1}^2 r_{5}^2 - r_{1}^2 r_{K}^2 - r_{5}^2
  r_{K}^2}].
\end{eqnarray}
Now when the one and five brane charges are equal, we choose $r_{a} \equiv R$,
$\tilde{\nu} \equiv \phi_{a}$ and $\tilde{\lambda} \equiv \sqrt{3}
\phi_{b}$. This implies $\sin{\psi} = 1$ and fixes the sign as being
positive in (\ref{hp}). 

The equations for the transformed fields (\ref{up}) have the same form
as the equation for the fixed scalar ${\nu}$, which was solved
first in \cite{CGKT} under certain conditions (extremality and low
energy very near extremality) and later in \cite{KK1} under more general
conditions. As usual there does not appear to be an analytic solution,
and we must patch together a solution between the near region (region
I, $r \ll r_{i}$), the intermediate region (region II, $r_{0} \ll r \ll
\omega^{-1}$) and the far region (region III, $r \gg r_{i}$). 

Following \cite{CGKT} and \cite{KK1}, we may write down the dominant
terms and the approximate solutions in the three regions as 
\begin{eqnarray}
{\rm I.} && [(h r^3 \partial_{r})^2 + r_{1}^2 r_{5}^2 (r^2 + r_{K}^2) \omega^2
- 8 r^4 h ] \phi^{I}_{a,b} = 0 \hspace{10mm} \phi_{a,b}^{I} = E
\frac{r^2}{r_{0}^2} + G; 
\nonumber \\
{\rm II.} && [(r^3 \partial_{r})^2 - 8 \frac{r_{a,b}^4}{(1 +
  \frac{r_{a,b}^2}{r^2})^2}] \phi^{II}_{a,b} = 0  \hspace{10mm}
\phi_{a,b}^{II} = \frac{C_{a,b}}{(1 + \frac{r_{a,b}^2}{r^2})} + D_{a,b} 
(1 + \frac{r_{a,b}^2}{r^2})^2;
\label{sol} \\
{\rm III.} && [(r^3 \partial_{r})^2 + r^6 \omega^2 ] \phi_{a,b}^{III} = 0  
\hspace{10mm} 
\phi_{a,b}^{III} = \alpha_{a,b} \frac{J_{1}(\omega r)}{\omega r} +
\beta_{a,b} \frac{ N_{1}(\omega r)}{\omega r}, \nonumber
\end{eqnarray}
where $C_{a,b}$, $D_{a,b}$, $\alpha_{a,b}$, $\beta_{a,b}$ are
constants. The full solution in the
inner region is obtained in terms of hypergeometric functions
\cite{KK1}, and we give only the limiting form for $r \rightarrow r_{0}$.
The quantity $E$ is fixed by the requirement that the solutions are purely
ingoing at the horizon to be 
\begin{equation}
E = \frac{2 \Gamma(1 - ia -ib)}{\Gamma(2-ia) \Gamma(2 -ib)},
\end{equation}
The quantity $G$ is similarly fixed, but we will not need it here. 
The constants $a$ and $b$ are
related to the effective left and right moving temperatures as 
\begin{equation}
a = \frac{\omega}{4 \pi T_{L}}, \hspace{10mm} b = \frac{\omega}{4 \pi
  T_{R}}.
\end{equation}
Matching between the three regions we need only retain the constants
\begin{equation}
\alpha_{a,b} = 2A_{a,b} = 2 E \frac{r_{a,b}^2}{r_{0}^2},
\end{equation}
provided that we calculate the absorption cross-sections by the ratio
of fluxes method \cite{MS}. 
The absorption probability for each scalar is given by the ratio of
the incoming fluxes at the horizon and at infinity where the flux of a
scalar field $f$ is given by 
\begin{equation}
F = \frac{1}{2i} (f^{\ast} h r^3 \partial_{r} f - c.c)
\end{equation}
and thus the absorption probability for the fixed scalar $\phi_{a}$ is 
\begin{equation}
P^{\phi_{a}}_{abs} = \frac{F_{horizon}}{F_{\infty}} = 2 \pi r_{1} r_{5}
\sqrt{r_{0}^2 + r_{K}^2} \omega^3 \frac{r_{0}^4}{4 |E|^2 r_{a}^4},
\end{equation}
and the absorption cross-section for $\phi_{a}$ is then given by 
\begin{equation}
\sigma_{abs}^{\phi_{a}} = \frac{9 \pi^3 r_{1}^6 r_{5}^6}{64 (r_{1}^2 +
  r_{5}^2 + \sqrt{r_{1}^4 + r_{5}^4 - r_{1}^2r_{5}^2})^2} 
\omega (\omega^2 + 16 \pi^2 T_{L}^2 ) 
  (\omega^2 + 16 \pi^2 T_{R}^2 ) \frac{e^{\frac{\omega}{T_{H}}}
  -1}{(e^{\frac{\omega}{T_{L}}} - 1)(e^{\frac{\omega}{T_{R}}} - 1)},
\end{equation}
which agrees with the result of \cite{KK1} when $r_{1} \equiv r_{5}$.
We have restricted to the dilute gas region $r_{0}^2 , r_{K}^2 \ll
r_{1}^2, r_{5}^2$ and accordingly dropped terms of order
$r_{0}^2/r_{1}^2$ and smaller. 
The corresponding result for the fixed scalar $\phi_{b}$ is 
\begin{equation}
\sigma_{abs}^{\phi_{b}} = \frac{9 \pi^3 r_{1}^6 r_{5}^6}{64 (r_{1}^2 +
  r_{5}^2 - \sqrt{r_{1}^4 + r_{5}^4 - r_{1}^2 r_{5}^2})^2} 
\omega (\omega^2 + 16 \pi^2 T_{L}^2 ) 
  (\omega^2 + 16 \pi^2 T_{R}^2 ) \frac{e^{\frac{\omega}{T_{H}}}
  -1}{(e^{\frac{\omega}{T_{L}}} - 1)(e^{\frac{\omega}{T_{R}}} - 1)},
\end{equation}
where we have again retained only leading order terms. Note that when
$r_{1} = r_{5}$ the absorption cross-section for $\lambda$ is a factor
of nine greater than that for $\nu$. 

We can use the solutions for the fixed scalars to find the gravitational
perturbations $\delta a$ and $\delta b$. Substitution into the two
remaining Einstein equations then provides a consistency check on our
results. The complexity of the graviton equations of motion implies
that this check is non-trivial to perform, but it may be verified that
the solutions obtained for the gravitational perturbations are indeed
consistent. 

\section{D-brane analysis}
\noindent

The string theory prediction of the absorption cross-section for the
fixed scalar $\nu$ was calculated in \cite{CGKT} where it was shown
that the semiclassical and string cross-sections agree when $r_{1} =
r_{5}$. Now to compare the semi-classical and string emission rates we should
calculate the string predictions for the cross-sections for the
scalars $\phi_{a,b}$. In terms of these scalars, the action (\ref{act}) 
takes the form
\begin{equation}
S_{5} = \frac{1}{2 \kappa_{5}^2} \int d^5x \sqrt{g} [R - 4 (\partial
\phi_{a})^2 - 4 (\partial \phi_{b})^2 - ..].
\end{equation}
From \cite{CGKT}, to study the leading order couplings of the fixed
scalars it is sufficient to retain the following terms in the effective
string action 
\begin{eqnarray}
I = \int d^2 \sigma \lbrace \frac{1}{2} (\partial_{+}X 
\partial_{-}X) + \frac{1}{4 T_{eff}} \nu (\partial_{+}X)^2
(\partial_{-}X)^2 \\
- \frac{1}{8 T_{eff}} \lambda [\partial_{+} X
\partial_{-} X ( (\partial_{+}X)^2 + (\partial_{-} X)^2 ) + 
(\partial_{+}X)^2 (\partial_{-}X)^2 ], \rbrace \nonumber 
\end{eqnarray}
where we have absorbed $\sqrt{T_{eff}}$ into the fields to make
them properly normalised. In terms of the fields $\phi_{a,b}$, the
relevant terms in the action are 
\begin{eqnarray}
I = \frac{1}{8 T_{eff}}\int d^2 \sigma \lbrace  
[\phi_{a} (2 \sin\psi - \sqrt{3} \cos\psi) + \phi_{b} (2 \cos\psi + \sqrt{3}
\sin\psi)](\partial_{+}X)^2 (\partial_{-}X)^2 \\
-  \sqrt{3} [\cos\psi \phi_{a} +  \sin\psi \phi_{b} ]
\partial_{+} X \partial_{-} X ( (\partial_{+}X)^2+ (\partial_{-} X)^2 ) 
\rbrace. \nonumber 
\end{eqnarray}
Let us consider the cross-section for the field $\phi_{a}$. The
effects of the interaction term involving the operator of dimension
$(2,2)$ were considered in \cite{CGKT} and we can hence write down the
contribution to the absorption cross-section from this interaction as 
\begin{eqnarray}
\sigma_{abs}^{\phi_{a} (1)} = \frac{\pi r_{1}^2 r_{5}^2}{1024 T_{eff}^2}
(2 \sin\psi - \sqrt{3}\cos\psi)^2 \
\omega (\omega^2 + 16 \pi^2 T_{L}^2 ) 
  (\omega^2 + 16 \pi^2 T_{R}^2 ) \times \\
 \frac{e^{\frac{\omega}{T_{H}}}
  -1}{(e^{\frac{\omega}{T_{L}}} - 1)(e^{\frac{\omega}{T_{R} }}
  - 1)}, \nonumber 
\end{eqnarray}
We then have to consider the contributions to the cross-section from
the other two interaction terms, operators of dimensions $(3,1)$ and
$(1,3)$ respectively. Suppose we consider first the interaction of the fixed scalar with one right and
three left movers. If $p_{1}$, $p_{2}$ and $p_{3}$ are the left moving
energies and $q_{1}$ is the right moving one, the matrix element among
properly normalised states is 
\begin{equation}
\sqrt{\frac{3}{2}} \cos\psi \frac{\kappa_{5}}{T_{eff}}
\sqrt{\frac{p_{1}p_{2}p_{3}q_{1}} {\omega}}.
\end{equation}
To compute the rate for the process $\phi_{a} \rightarrow L + L + L +
R$ we have to square the normalised matrix element and integrate it
over the possible energies of the final state particles. Because of
the presence of the thermal sea of right and left movers we must
insert Bose enhancement factors; for example, each left mover in the
final state picks up a factor of $-\rho_{L}(-p_{i})$ where
\begin{equation}
\rho_{L} (p_{i}) = \frac{1}{e^{\frac{p_{i}}{T_{L}}} - 1} 
\end{equation}
is the Bose Einstein distribution. Similarly if there is a left mover in the
initial state, it picks up an enhancement factor
$\rho_{L}(p_{i})$.  
Conservation of energy and of momentum parallel to the effective
string then introduces the factor
\begin{equation}
(2 \pi)^2 \delta (p_{1} + p_{2} + p_{3} + q_{1} - \omega) 
\delta(p_{1} + p_{2} + p_{3} - q_{1}) = \frac{1}{2} (2 \pi)^2 \delta (
p_{1} + p_{2} + p_{3} - \frac{\omega}{2}) \delta (q_{1} -
\frac{\omega}{2}).
\end{equation}
To get the full rate, we should calculate first the 
rate for $\phi_{a} \rightarrow L + L + L + R$, and then include the
rates for other absorption processes such as $\phi_{a} + L \rightarrow
L + L + R$. However it is sufficient for our purposes to obtain the
functional dependence of absorption rate which following the
methods of \cite{CGKT} is given by
\begin{equation}
\Gamma \sim \frac{\kappa_{5}^2 L_{eff} \cos^2 \psi}{T_{eff}^2
  \omega} \int^{\infty}_{-\infty} dp_{1} dp_{2} dp_{3} \delta(p_{1} + p_{2} +
  p_{3} - \frac{\omega}{2}) \prod_{i=1,3} \frac{p_{i}}{1 -
  e^{-\frac{p_{i}}{T_{L}}} } \times \frac{\omega}{2( 1 - e^{-
  \frac{\omega}{2 T_{R}}})},
\end{equation}
where $L_{eff}$ is the length of the effective string given by
\begin{equation}
\kappa_{5}^2 L_{eff} = 4 \pi^2 r_{1}^2 r_{5}^2. 
\end{equation}
The prefactors
will be determined by the number of species of left and right movers
and symmetry factors. Integrating over the left movers we find that
the functional dependence of the emission rate is 
\begin{equation}
\Gamma \sim \frac{\kappa_{5}^2 L_{eff} \cos^2 \psi}{T_{eff}^2 } \frac{\omega}
 {(1 - e^{-\frac{\omega}{2T_{L}}}) (1 - e^{-\frac{\omega}{2T_{R}}})} 
(\omega^2 + 16 \pi^2 T_{L}^2) (\omega^2 + 32 \pi^2 T_{L}^2),
\end{equation}
and we obtain the same functional dependence for processes with one
left and three right movers with $T_{L}$ and $T_{R}$ exchanged. Now
the absorption cross-section is given by
\begin{equation}
\sigma_{abs} = \Gamma(\omega) + \Gamma(-\omega),
\end{equation}
and thus the contribution to the $\phi_{a}$ cross-section from the
$(1,3)$ and $(3,1)$ operators takes the form 
\begin{eqnarray}
\sigma_{abs}^{\phi_{a}(2)} & \sim &  \frac{\kappa_{5}^2
  L_{eff} \cos^2 \psi}{T_{eff}^2} \frac{\omega (e^{\frac{\omega}{T_{H}}} - 1)}
 {(e^{\frac{\omega}{2T_{L}}} - 1) (e^{\frac{\omega}{2T_{R}}} -1)}
 \times \\
&& \lbrace (\omega^2 + 16 \pi^2 T_{L}^2)(\omega^2 + 32 \pi^2 T_{L}^2)
  + (\omega^2 + 16 \pi^2 T_{R}^2)(\omega^2 + 32 \pi^2
  T_{R}^2) \rbrace. \nonumber
\end{eqnarray}
The cross-section for the other scalar takes a similar form with
appropriate normalisation factors. 
Thus whatever the value of $r_{1}$ and $r_{5}$ there is no agreement
between the semi-classical and string cross-sections for the fixed
scalars. Even for $r_{1} = r_{5}$, when the cross-sections for the
fixed scalar $\phi_{a} = \nu$ are in agreement, the functional
dependences of semi-classical and string 
cross-sections for the other fixed scalar differ. 

For general charges, if we consider low energy absorption by a near
extremal black hole, the temperature of the left-movers is much
greater than that of the right-movers, and the dominant term in the
string cross-section for $\phi_{a}$ is 
\begin{equation}
\sigma_{abs}^{\phi_{a}} \rightarrow \frac{r_{1}^2 r_{5}^2 \cos^2 \psi}
{T_{eff}^2} T_{L}^5,
\end{equation}
whereas the semi-classical cross-section behaves as 
\begin{equation}
\sigma_{abs}^{\phi_{a}} \rightarrow \frac{r_{1}^6r_{5}^6}{(r_{1}^2 +
  r_{5}^2 + \sqrt{r_{1}^4 + r_{5}^4 - r_{1}^2 r_{5}^2})^2} \omega^2 T_{L}^3, 
\end{equation}
and thus vanishes as the frequency goes to zero. 

\section{Comparison of greybody factors}
\noindent

We have found that for general charges the functional dependences of
the string and semi-classical absorption cross-sections for the fixed
scalars differ and that there is a mixing between the two scalars in
the semi-classical equations which is not explained by the effective string 
model. It is interesting to notice that the functional dependences of
both semi-classical cross-sections are determined by the behaviour in the near
horizon region. At very small radius, one can see from the equations
for the fixed scalars that both $\lambda$ and $\nu$ satisfy
hypergeometric equations. Mixing between gravitational and fixed
scalar perturbations in the intermediate regions determines the
normalisations of the cross-sections. 
 
The results for the semi-classical absorption cross-sections of the
fixed scalars imply that the cross-sections for
both have the same functional dependence on the energy and the left
and right moving temperatures. Since the fixed scalar $\nu$ couples
only to an operator of dimension $(2,2)$ in the effective string
action, this suggests that the fixed scalar $\lambda$ couples in a
similar way. However, the expression for the string action 
includes couplings to operators of dimensions $(3,1)$ and $(1,3)$ and
thus the semiclassical and string cross-sections disagree. 

There are several possible explanations for this discrepancy. If we
assume that the semi-classical calculation is correct, agreement might
be restored by the presence of additional couplings within the string
effective action. Indeed, since fermions had to be included in 
\cite{CGKT} in order for the normalisations of cross-sections to agree
when $r_{1} = r_{5}$, it seems feasible that further modifications of
the effective action may be required. One can calculate the form of
the effective string action needed to give agreement between string
and semi-classical calculations as
\begin{eqnarray}
I = \int d^2 \sigma [\frac{3}{4 T_{eff}} \phi_{a} (\frac{r_{1}^2}{(
  r_{1}^2 + r_{5}^2 + \sqrt{r_{1}^4 + r_{5}^4 - r_{1}^2 r_{5}^2})})
  (\partial_{+} X)^2 (\partial_{-} X)^2  \\
 + \frac{3}{4 T_{eff}} \phi_{b} (\frac{r_{1}^2}{(
  r_{1}^2 + r_{5}^2 - \sqrt{r_{1}^4 + r_{5}^4 - r_{1}^2 r_{5}^2})})
  (\partial_{+} X)^2 (\partial_{-} X)^2 ],
\end{eqnarray}
where we have assumed $T_{eff} = 1/ 2 \pi r_{5}^2$. 
Thus the normalisation of the interaction terms in the effective string action
would have to depend on the one and five brane charges and such a
dependence is difficult to explain. 
The importance of resolving this disagreement need not be stressed
given its relationship both to the information loss question and
to further understanding of the effective string model. 

\bigskip

{\bf Note:} Whilst this work was being completed, reference 
\cite{KK2}, which has considerable overlap with this paper, appeared.

\end{document}